\documentclass[aps,pre,preprint,showpacs]{revtex4}
\usepackage{graphicx}
\usepackage{latexsym}
\usepackage{color}

\begin{document}

\title{An instrumented tracer for Lagrangian measurements in Rayleigh-B\'{e}nard convection}

\author{Woodrow L. Shew, Yoann Gasteuil, Mathieu Gibert, Pascal Metz \& Jean-Fran\c{c}ois Pinton}
\affiliation{Laboratoire de Physique de l'\'{E}cole Normale
Sup\'{e}rieure de Lyon, CNRS UMR5672, \\
46 all\'ee d'Italie F-69007 Lyon, France}

\begin{abstract}
We have developed novel instrumentation for making Lagrangian
measurements of temperature in diverse fluid flows. A small
neutrally buoyant capsule is equipped with on-board electronics
which measure temperature and transmit the data via a wireless
radio frequency link to a desktop computer.  The device has 80 dB
dynamic range, resolving milli-Kelvin changes in temperature with
up to 100 ms sampling time.  The capabilities of these ``smart
particles" are demonstrated in turbulent thermal convection in
water.  We measure temperature variations as the particle is
advected by the convective motion, and analyse its statistics.
Additional use of cameras allow us to track the particle position
and to report here the first direct measurement of Lagrangian heat
flux transfer in Rayleigh-B\'{e}nard convection.  The device shows
promise for opening new research in a broad variety of fluid
systems.
\end{abstract}


\maketitle

\section{Introduction}
\label{intro} Scalar mixing in turbulent flows plays a crucial
role in uncountable natural, medical, and industrial systems:
spread of pollutants by wind and water; oceanic and atmospheric
thermal convection; the life cycles of plankton;  mixing in
combustion engines and chemical reactors.  A natural approach to
understanding these examples is to measure the trajectories of the
pollutant molecules, warm fluid elements, planktonic organisms,
and chemical species, respectively, as well as the properties of
the flow along these trajectories.  This Lagrangian approach to
turbulent mixing has been advanced significantly with numerical
and theoretical models \cite{GawedzkyReview}, but experimental
works are rare due to the difficulties in (1) distinguishing the
identity of a fluid particle along its trajectory, and (2) making
measurements along these paths . A few recent experimental studies
have successfully performed resolved measurements of the
trajectories of small solid tracer particles in turbulence
\cite{Riso,pinton,bodenschatz,ETH}. Atmospheric research
groups~\cite{ballons} and oceanographers routinely employ
meter-sized Lagrangian floats to study large scale
flows~\cite{lagfloat}. Lagrangian probes have a distinct advantage
over Eulerian probes which record at fixed positions in the flow:
the measurements provide information about local physical
processes as experienced by the fluid particle.

With the aim of measuring Lagrangian scalar quantities in
well-controlled laboratory flows, we have developed a miniature,
wireless, neutrally-buoyant instrument, dubbed a {\it smart
particle}.  We present temperature measurements obtained with this
device in turbulent thermal convection.  Similar instruments have
been used in medical applications~\cite{medical}, but to our
knowledge we report the first such measurement device in the 
context of fluid dynamics. Simultaneous measurements of
temperature and position have been made before in convection
experiments, most recently by tracking optically the motion of
microspheres containing thermochromic liquid crystal
(TLC)~\cite{Zhou}. The shape evolution of thermal plumes very
close to the thermal and hydrodynamic boundary layers was
quantified, but this method is not capable of tracking individual
particles over large distances. Our smart particle measurements
complement those of TLC particles by vastly improving temperature
and time resolution as well as allowing observation of very long
particle trajectories, though the smart particles are too large to
investigate the flow in the boundary layers.

\section{Device details}\label{dev}
The details of the instrument are describe in the following five
subsections. First we present a block diagram and general
description of the whole system. Then, the radio frequency (RF)
data transfer is described in detail.   Next, we focus on the
power management system. We then discuss the measurement and
processing of temperature data and end with details about the
position measurement technique.

\subsection{Overview}
The smart particle consists of a $D = 21$~mm diameter capsule
containing temperature instrumentation, an RF emitter, a battery,
and an on-off switch -- illustrated in Figs.~1(a,b).  A
resistance controlled oscillator is used to create a square wave
whose frequency depends on the temperature of several thermistors.
This square wave is used directly to modulate the amplitude of the
radio wave generated by the RF emitter.  The entire mobile circuit
is powered with a coin cell battery and may be put in a low power
standby mode using an externally applied magnetic field.

The stationary parts of the system include an antenna, an RF
receiver, 2 RF amplifiers, a high speed data acquisition system,
and a desktop computer running Labview (Fig.~1a).  The
broad-band radio frequency amplifiers increase the voltage
amplitude of the signal by a factor of 26~dB. The receiver is
carefully tuned to demodulate the signal produced by the emitter.
The receiver outputs a square wave identical to that generated by
the resistance controlled oscillator.  The frequency of this
square wave, and hence temperature, is recovered on-the-fly using
a Labview algorithm.  In addition, the particle trajectory is
recorded with a digital video camera, resulting in synchronous
measurements of the position and temperature of the particle as it
is carried about by the fluid.

\subsection{RF data link}
The radio frequency emitter is a MAX7044 (from Maxim Integrated
Products), which employs on-off keying (OOK) amplitude modulation.
The RF carrier frequency is 315 MHz. The MAX7044 expects a digital
modulating signal with a baud rate from 0 to 100 kHz. We implement
an 8~mm diameter, 8-turn coil emitter antenna with about 200 nH
inductance and 4~$\Omega$ resistance. We use a split-capacitor
impedance matching network to maximize the efficiency since the
MAX7044 is optimized for a 125~$\Omega$ antenna.  The emitter
draws an average of 5 mA and requires 2.1-3.6 V from the battery
in normal operation.  The emitter consumes 10 to 100 times more
power than any of the other on-board components.

On the receiving end we use the MAX1473 superheterodyne receiver,
which demodulates OOK data and is designed to work with the
MAX7044. The receiving antenna is a 315 MHz 1/4 wave whip (Linx
Technologies, Inc., ANT-315-CW-HD). Between the antenna and the
receiver are two low noise amplifiers each with a gain of 13 dB at
315 MHz (MAX2640).

\subsection{Power management}
The size of the instrumentation capsule is minimized so that it
capable of probing as small a spatial scale as possible. The
component of the system which limits the size of the capsule most
is the battery.  It is a CR1616 lithium coin cell (Panasonic),
which is 16 mm in diameter and 1.6 mm thick. This battery can supply the necessary power for about 3~hours. The circuit is also equipped
with a magnetic field triggered switch so that the instrumentation
may be turned off when we are not ready to acquire data.  The
switch is composed of a TLE4913 low power hall switch (Infineon
Technologies) and a NL17SZ74 single D flip-flop (ON
Seminconductor).  The hall switch outputs a high logic level when
in the presence of a sufficiently large magnetic field. Feeding
this output into the CP pin of the flip-flop while D is connected
to $\bar{{\rm Q}}$ we have a ``push-button" switch, which is
operable with a hand-held permanent magnet from a distance of
several centimeters. When the switch is in the ``off" position,
the circuit consumes less than 100 $\mu$A.  The hall switch and
the flip-flop respectively require 2.4-5.5 V and 1.65-5.5 V from
the battery.

\subsection{Temperature detection}
The system measures the spatially averaged temperature around the
smart particle body.  This is accomplished using four thermistors
to set the frequency of a LMC555 timer/oscillator (National
Semiconductor) operating in astable mode.  The thermistors
protrude from the capsule wall by about 0.5 mm and are spaced
evenly around its middle.  The LMC555 outputs a square wave at
logic levels with a period $1/f=(R_1+R_2+2R_3+2R_4)C/1.44$, where
$R_i$ are the resistances of the four thermistors, and C is 47~pF.
We employ 0.8~mm, 230~k$\Omega$ thermistors (Epcos B57540G0234)
with a response time of about 0.06~s in water, which is faster
than the fastest temperature time scales of the flow we study.  In
the range of temperature $26-34~^\circ$C, the frequency of the
square wave is in the range $22-26$ kHz; the sensitivity is
513~Hz/$^\circ$C. The relationship between temperature and
frequency is linear within 2 percent. The LMC555 requires about
100~$\mu$A at a supply voltage between 1.5 and 12~V.

The RF receiver recovers the square wave signal, which is then
recorded with a high-speed, 14 bit, analog-to-digital converter
(ADC) (National Instruments 5621 in a PXI system).  Running on a
desktop PC, Labview is used to control the data acquisition. The
frequency of the demodulated square wave is computed on-the-fly
using a standard Labview library.  Each measurement of frequency
is computed from 100~ms of the square wave, i.e. about 3000
periods, sampled by the ADC at 10 MHz.  The resulting measurement
resolution is about $\pm2$~Hz, which is $\pm4$ mK when converted
to temperature.  The sampling rate is about 10~Hz.  With maximum
flow velocities in the range 1-2~cm/s and a particle size of
21~mm, we are oversampling the dynamics by a factor of order 10.

\subsection{Position measurement}
In order to measure the smart particle position, we use a standard
webcam, interfaced with a desktop computer.  A uniform
well-illuminated image background is achieved with a back-lit
sheet of frosted glass.  We use Matlab scripts to control the
camera and process the video data.  First, each video frame is
converted into a 2D binary array, using an adaptive threshold.
Then, the particle position is extracted using an image
recognition algorithm.  The effective sample frequency is around
5~Hz. The resolution of the camera is 640 $\times$ 480 pixels, so
that the particle position is determined with 0.1~mm precision.

\section{Measurements in Rayleigh-B\'enard convection}

\subsection{Convection apparatus}
Our experimental setup, shown in figure~1c, is a
traditional rectangular Rayleigh-B\'enard cell with height
$H=40$~cm and section $40~$cm~$\times$~10~cm. The fluid is water
and the walls of the vessel are 25~mm thick PMMA. The upper plate
of the cell is temperature regulated by a chilled water bath. The
bottom plate is heated by 5 resistors, regularly spaced. Complete
experimental details can be found in~\cite{Chilla}.

In the results reported here, the power input is $P=230$~W, and
the upper plate is regulated to $T_{\rm up}=19^\circ$C,
corresponding to a temperature difference between the top and
bottom plates of $\Delta T = 20.3^\circ$C. As a result, the
Rayleigh number is
\begin{equation}
Ra = \frac{g\beta \Delta T H^3}{\nu \kappa} = 3.07\times10^{10},
\end{equation}
where $g$ is acceleration due to gravity, $\beta = 2.95 \times
10^{-4} \; {\rm K}^{-1}$ is the thermal
expansion coefficient of water and $\nu = 8.17 \times 10^{-7}{\rm
m}^{2}{\rm s}^{-1}$, $\kappa = 1.48 \times 10^{-7}{\rm m}^{2}{\rm
s}^{-1}$ its viscosity and thermal diffusivity (values are given
for a mean temperature equal to 29.1$^\circ$C). The Nusselt
number, measured as the total heat flux normalized by
$\kappa\Delta T/L$, is $Nu = 167.9 \pm 0.2$. Under these
conditions, the convective regime is fully turbulent~\cite{Chilla,
Castaing} and the mean flow is a steady, system-sized, single
convection roll with a rotation period of about 100~s.  The
characteristic thickness of the thermal boundary layer is $\ell_T
\sim \frac{1}{2} H Nu^{-1} \sim 1.2$~mm and that of the
hydrodynamic boundary layer is $\ell_U \sim
\ell_T(\nu/\kappa)^{1/3} \sim 2$~mm.  Thus, the particle is too
large to penetrate the boundary layers.

\subsection{Temperature measurement}
The particle and fluid density are carefully matched. This is
achieved by initially adjusting the particle mass to within 1
percent of $\rho_f \pi D^3/6$, where $\rho_f$ is the density of
water. Then, to obtain neutral buoyancy within 0.05 percent,
$\rho_f$ is finely adjusted by the addition of small amounts
(about 1000 ppm) of pure glycerol, which has a density 20 percent
greater than water.  The particle is inevitably slightly lighter
than the cool fluid near the upper plate and slightly heavier than
the warm fluid near the bottom plate. Nonetheless, the
particle explores all regions of the vessel without a clear bias
as can be seen in figure~2b.

We show in figure~3a, a time series of the
temperature $\theta (t)$ recorded by the particle, during the
trajectory shown in figures~2b,c.  Apparent in the time
series is a nearly periodic, large amplitude fluctuation, which
reflects the particle's entrainment in the mean flow.  This
periodicity, as well as the smaller amplitude turbulent
fluctuations are made more clear in the power spectrum shown in
figure~3b.  The spectrum also reveals a range of
time scales, corresponding to frequencies $0.016 < f < 0.230$~Hz,
which is consistent with a self-similar $\tilde{\theta}^2 \sim
f^{-\alpha}$ scaling, with an exponent $\alpha$ close to -2. This
feature is in sharp contrast with the Eulerian temperature
spectrum obtained when the particle is constrained to remain at
fixed location in space.  In that case, we have verified that the
spectrum shows an $f^{-7/5}$ scaling range. This value is in
agreement with other experimental Eulerian studies of convection
(e.g. \cite{7over5}), provided the frequency spectrum be
interpreted as a spectrum in space, which is plausible considering
the persistent mean flow (i.e. Taylor's hypothesis).
Bolgiano-Obhukov scaling also leads to a $k^{-7/5}$ wavenumber
spectrum for temperature variations in real space. By contrast,
the spectral behavior found here for the moving particle is
reminiscent of velocity data for Lagrangian tracers in turbulent
flows. Indeed when fluid turbulence is fully developed, a
well-established feature of Lagrangian dynamics and Kolmogorov's
theory is that the Lagrangian velocity spectrum should have an
$f^{-2}$ inertial range~\cite{pinton,yeung} -- as we have verified also verified here from our independent measurement of the particle position. The upper cut-off
frequency apparent in figure~3b around ($\sim
0.3$~Hz) corresponds to the characteristic time of motion of the
particle across a distance equal to its diameter $D$: the mean
speed of the large scale roll being $\approx 1.6~ {\rm cm}{\rm
s}^{-1}$. This is also in agreement with Lagrangian measurements
made using density matched tracers: the velocity spectrum was
found to follow the expected Lagrangian behavior up to a frequency
set be the particle's size and the flow characteristic velocity
scale \cite{pinton}.  These interpretations are consistent with
qualitative observations using Schlieren visualization, which
reveal that temperature plumes of size $D$ or larger fully entrain
the smart particle, while smaller plumes do so only partially.

Finally, as shown in figure~3c, the histogram of
temperature is non-Gaussian.  The non-Gaussian character of the
PDF is in agreement with Eulerian measurements using fixed
temperature probes~\cite{Castaing}.  Other recent experimental
studies have also shown that the distribution of intense
temperature plumes in the form of mushrooms has non Gaussian
tails~\cite{Zhou}.  We note that the asymmetry of the PDF is
probably due to imperfect density matching between the fluid and
the particle, so that it tends to stay slightly longer near the
bottom plate.

\subsection{Heat flux measurement}
With the assumption that the particle may indeed act as a
Lagrangian tracer in the convective flow, it is natural and
interesting to discuss the heat carried by the flow in terms of
the temperature measured by the particle. To this end, we define
an instantaneous Lagrangian Nusselt number as:
\begin{equation}
Nu^{L}(t) = 1 + \frac{L}{\kappa \Delta T} \theta'(t) \cdot v_z(t) \ ,
\end{equation}
where $\theta'(t) = \theta(t) - \overline{\theta}$ is the particle
temperature variation from its time averaged temperature
$\overline{\theta}$, and $v_z(t)$ its vertical velocity.
Integrating this quantity over all the fluid particles in any
horizontal plane recovers the traditional Nusselt number.

The time series of $Nu^{L}$ is shown in figure~4a.
Comparing this time series with the trajectory in figure~2(c), we
see that the heat transfer intermittently spikes to very large
values often as the particle moves away from the end plates and
mixes with fluid at intermediate heights in the vessel.  The power
spectrum in figure~4b shows that the periodicity of the
large scale convection roll is diminished and strong fluctuations
occur over a broader range of (slow) scales. The spectrum shows a
fast, $f^{-4}$ decrease in the same frequency range that the
temperature showed $f^{-2}$.  As discussed in the previous
section, this scaling is consistent with our understanding of
velocity spectra ($\tilde{v}^2\sim f^{-2}$) resulting in
$\tilde{Nu^{L}}^2\sim \tilde{\theta}^2 \tilde{v}^2 \sim f^{-4}$.
The PDF of $Nu^{L}$, figure~4c is skewed toward positive
values, as expected because coherent convective motion are
associated with either hot fluid rising or cold fluid sinking, so
that in each case $\theta' \cdot v_z >0$. The much less probable
events with $\theta' \cdot v_z <0$ correspond, for instance, to
the rise of the particle when it is colder than its environment.
This is unlikely but not impossible as the particle may be trapped
in the swirls of a turbulent plume (e.g. the flow in the ``cap" of
the mushroom shaped plume is opposite the flow in the ``stem").
The most probable value of the PDF is zero, indicating the
particle spends a large amount of time at the mean fluid
temperature in the bulk of the cell. The mean value,
$\overline{Nu^L}$, is more surprising: at 335.4 it is roughly
twice the global, Eulerian, Nusselt value computed from power
input and velocity differences. Such an increased mean may be due
to the fact that the particle does not sample the flow uniformly,
but may instead get more advection from intense thermal plumes.
The dramatically non-Gaussian form of the PDF as well as
scale-by-scale statistics of the local flux requires further
study. This analysis will be reported elsewhere.

\section{Conclusions}
We report on new instrumentation for making fluid mechanics
measurements in the reference frame of particles passively
advected by flow motion.  These smart particles show promise for a
fresh perspective in turbulence research as well as two phase
flows. The rapid development of MEMS and microfluidic components,
which may be incorporated in a smart particle, opens a diverse
range of potential applications including granular flows and
reacting flows in chemical and biological systems.

We demonstrate the capabilities of smart particles in the 
simple case of Lagrangian temperature measurements in thermal
convection.  We report several original findings. First, we
observe $f^{-2}$ power law scaling behavior for the temperature
spectrum, which suggests the temperature acts as a passive scalar
advected by fully developed, Kolmogorov-like, turbulence.  This
result supports the hypothesis that outside the boundary layers,
where buoyancy sets the fluid into motion, the temperature is
passively mixed in the turbulent bulk. (e.g. \cite{ss, Castaing2,
zhou2}). Strong temporal intermittency in Lagrangian heat
transport is revealed in the extremely heavy-tailed probability
distribution functions, while the fact that the mean Lagrangian
heat transport is much higher than the Eulerian average suggests
strong spatial inhomogeneity as well. Further and deeper
investigations of these features will be presented soon elsewhere.\\

\noindent{\bf Acknowledgements}\\
The authors have had fruitful discussions with Francesca Chill\'a and Bernard Castaing. This work has been supported by C.N.R.S. and Emergence Contract No.~2005-12 from the French Rh\^one-Alpes Region.
 

\newpage\clearpage





\begin{figure}
\centerline{\includegraphics[width=17cm]{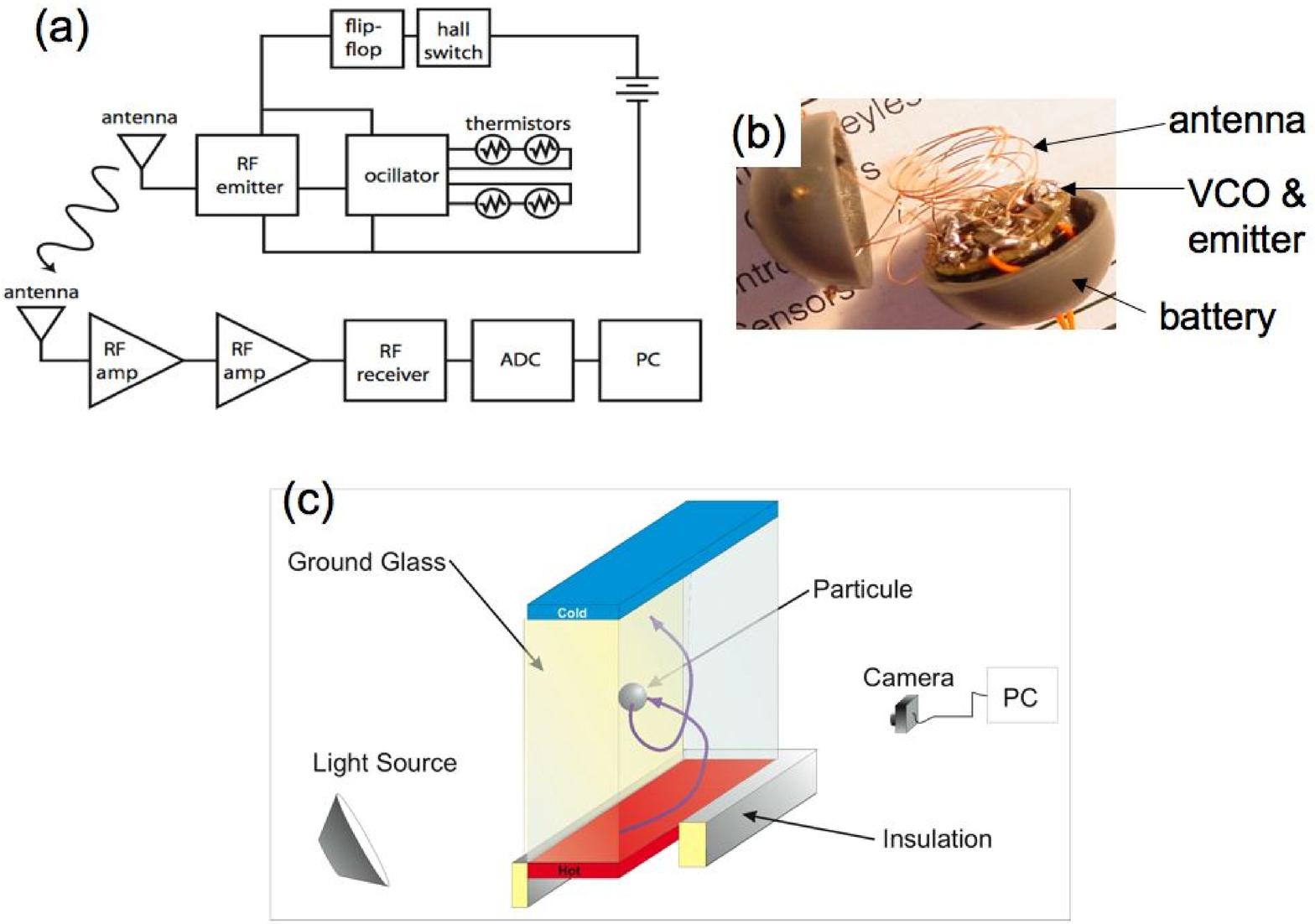}}
\caption{ (a) Block diagram of measurement system. (b) Photo of the
smart particle before closing the capsule.} \label{block}
\end{figure}

\newpage\clearpage
\begin{figure}
\centerline{\includegraphics[width=17cm]{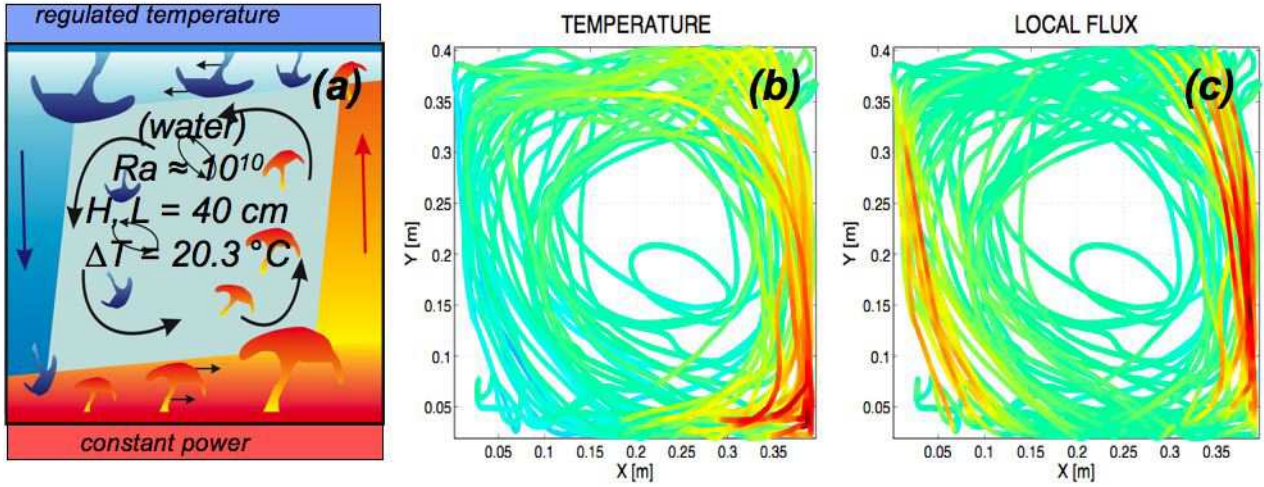}}
\caption{(a) Cartoon of the convection flow. The black disc
represents the smart particle. (b,c) Trajectory of the particle
with temperature (b) and heat flux (c) encoded in the color. }
\label{cell}
\end{figure}

\newpage\clearpage
\begin{figure}[t!]
\centerline{\includegraphics[width=17cm]{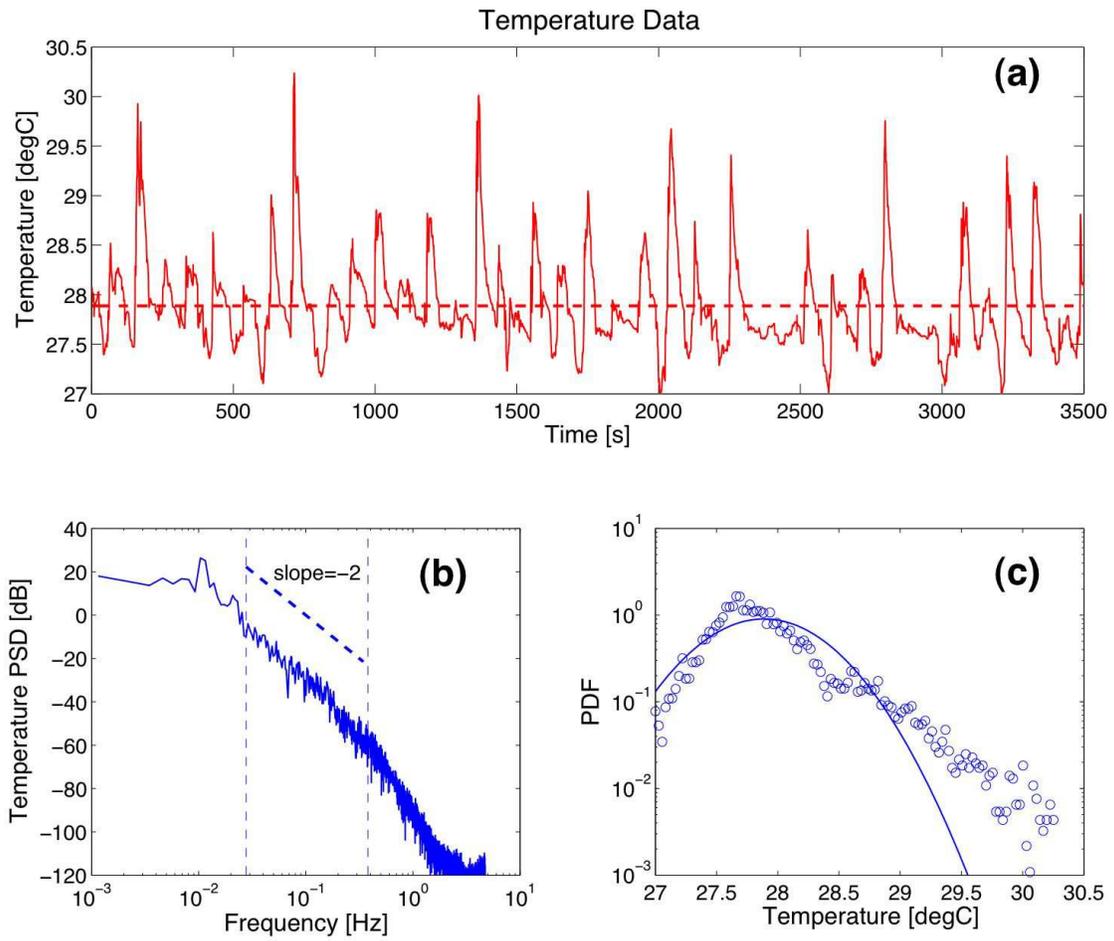}}
\caption{Temperature measurements: (a) time series for the trajectory in figure~2(a). The mean temperature is $27.887$~degC and the standard deviation is $0.451$~degC; (b) corresponding power spectrum; (c) Temperature probability density function (circles), compared to a Gaussian (solid line).}
\label{temperature}
\end{figure}

\newpage\clearpage
\begin{figure}[h!]
\centerline{\includegraphics[width=17cm]{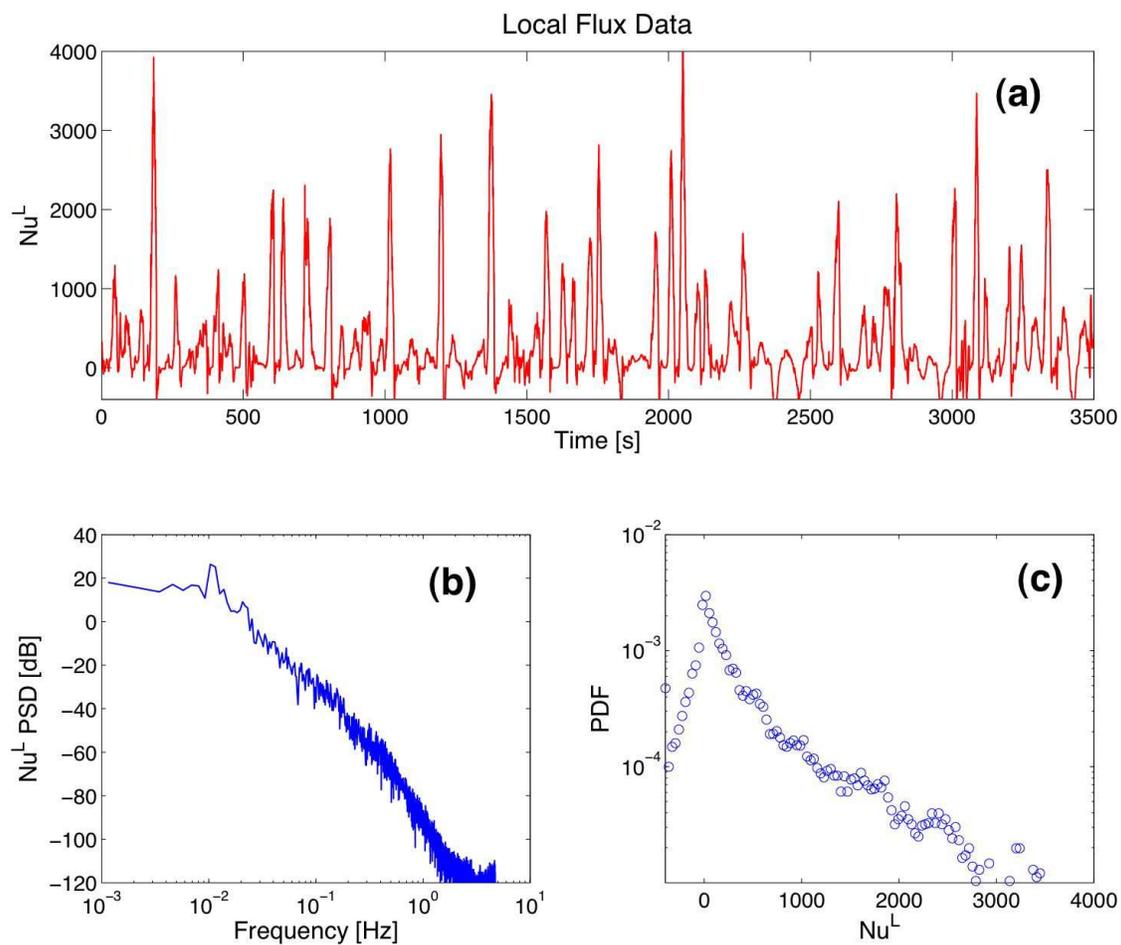}}
\caption{Heat flux measurements: (a) time series of $Nu^{L}(t)$ for the trajectory in figure~2(a); (b) corresponding power spectrum; (c) Probability density function.} \label{flux}
\end{figure}

\end{document}